\documentclass[10pt,conference]{IEEEtran}
\usepackage{amsmath,graphicx,url}
\usepackage{multirow}
\usepackage{todonotes}
\usepackage{amsfonts}
\IEEEoverridecommandlockouts
\title{Direction of Arrival Estimation \\ for Multiple Sound Sources Using \\Convolutional Recurrent Neural Network}
\author{\IEEEauthorblockN{Sharath Adavanne\textsuperscript{*}\textsuperscript{1}, Archontis Politis\textsuperscript{*}\textsuperscript{2}, Tuomas Virtanen\textsuperscript{1} \thanks{\textsuperscript{*}Equally contributing authors in this paper. The research leading to these results has received funding from the European Research Council under the European Union’s H2020 Framework Programme through ERC Grant Agreement 637422 EVERYSOUND. The authors also wish to acknowledge CSC-IT Center for Science, Finland, for computational resources}}
\IEEEauthorblockA{\textsuperscript{1}Laboratory of Signal Processing, Tampere University of Technology, Finland\\
\textsuperscript{2}Department of Signal Processing and Acoustics, Aalto University, Finland }
}

\begin{document}
\maketitle
\bstctlcite{IEEEexample:BSTcontrol}
\begin{abstract}
This paper proposes a deep neural network for estimating the directions of arrival (DOA) of multiple sound sources. The proposed stacked convolutional and recurrent neural network (DOAnet) generates a spatial pseudo-spectrum (SPS) along with the DOA estimates in both azimuth and elevation. We avoid any explicit feature extraction step by using the magnitudes and phases of the spectrograms of all the channels as input to the network. The proposed DOAnet is evaluated by estimating the DOAs of multiple concurrently present sources in anechoic, matched and unmatched reverberant conditions. The results show that the proposed DOAnet is capable of estimating the number of sources and their respective DOAs with good precision and generate SPS with high signal-to-noise ratio.
\end{abstract}

\section{Introduction}
\label{sec:intro}
Direction of arrival (DOA) estimation is the task of identifying the relative position of the sound sources with respect to the microphone. DOA estimation is a fundamental operation in microphone array processing and forms an integral part of speech enhancement~\cite{Woelfel2009}, multichannel sound source separation \cite{nikunen2014direction} and spatial audio coding~\cite{politis2015sector}. Popular approaches to DOA estimation are based on time-delay-of-arrival (TDOA)~\cite{Huang2001}, the steered-response-power (SRP) ~\cite{Brandstein1997}, or on subspace methods such as multiple signal classification (MUSIC)~\cite{Schmidt1986} and the estimation of signal parameters via rotational invariance technique (ESPRIT)~\cite{Roy1989}. 

The aforementioned methods differ from each other in terms of algorithmic complexity, and their suitability to various arrays and sound scenarios. MUSIC specifically is very generic with regards to array geometry, directional properties and can handle multiple simultaneously active narrowband sources. On the other hand, MUSIC and subspace methods in general, require a good estimate of the number of active sources, which are often unavailable or difficult to obtain. Furthermore, MUSIC can suffer at low signal to noise ratio (SNR) and in reverberant scenarios~\cite{dibiase2001robust}. In this paper, we propose to overcome the above shortcomings with a deep neural network (DNN) method, referred to as DOAnet, that learns the number of sources from the input data, generates high precision DOA estimates and is robust to reverberation. The proposed DOAnet also generates a spatial acoustic activity map similar to the MUSIC pseudo-spectrum (SPS) as an intermediate output. The SPS has numerous applications that rely on a directional map of acoustic activity such as soundfield visualizations~\cite{o2008imaging}, and room acoustics analysis~\cite{khaykin2012acoustic}. In comparison, the proposed DOAnet outputs the SPS and DOA's of multiple overlapping sources similar to any popular DOA estimators like MUSIC, ESPRIT or SRP without requiring the critical information of the number of active sound sources. A successful implementation of this will enable the integration of such DNN methods to higher-level learning based end-to-end sound analysis and detection systems.

Recently, several DNN-based approaches have been proposed for DOA estimation~\cite{Roden2015, Xiao2015, Takeda2016,Zermini2016, Vesperini2016, Chakrabarty2017}. There are six significant differences between them and the proposed method: a) All the aforementioned works focused on azimuth estimation, with the exception of ~\cite{Vesperini2016} where the 2-D Cartesian coordinates of sound sources in a room were predicted, and~\cite{Roden2015} trained separate networks for azimuth and elevation estimation. In contrast, we demonstrate the estimation of both azimuth and elevation for the DOA by sampling the unit sphere uniformly and predicting the probability of sound source at each direction. b) The past works focused on the estimation of a single DOA at every time frame, with the exception of~\cite{Takeda2016} where localization of azimuth for up to two sources simultaneously was proposed. On the other hand, the proposed DOAnet does not algorithmically limit the number of directions to be estimated, i.e., with a higher number of audio channels input, the DOAnet can potentially estimate a larger number of sound events.

c) Past works were evaluated with different array geometries making comparison difficult. Although the DOAnet can be applied to any array geometry, we evaluate the method using real spherical harmonic input signals, which is an emerging popular spatial audio format under the name Ambisonics. Microphone signals from various arrays, such as spherical, circular, planar or volumetric, can be transformed to Ambisonic signals by an appropriate transform~\cite{teutsch2007modal}, resulting in a common representation of the 3-D sound recording. Although the DOAnet is scalable to higher-order Ambisonics, in this paper we evaluate it using the compact four-channel first-order Ambisonics (FOA).


d) Regarding classifiers, earlier methods have used fully connected (FC) neural networks~\cite{Roden2015, Xiao2015, Takeda2016,Zermini2016, Vesperini2016} and convolutional neural networks (CNN)~\cite{Chakrabarty2017}. In this work, along with the CNNs we use recurrent neural network (RNN) layers. The usage of RNN allows the network to learn long-term temporal information. Such an architecture is referred to as a convolutional recurrent neural network (CRNN) in literature and is the state-of-the-art method in many single-~\cite{sainath2015,Malik2017} and multichannel~\cite{Sainath2017, Adavanne2017} audio tasks. e) Previous methods used inter-channel features such as generalized cross-correlation with phase transform (GCC-PHAT)~\cite{Vesperini2016,Xiao2015}, eigen-decomposition of the spatial covariance matrix~\cite{Takeda2016}, inter-channel time delay (ITD) and inter-channel level differences (ILD)~\cite{Roden2015,Zermini2016}. More recently, Chakrabarty et al.~\cite{Chakrabarty2017} proposed to use only the phase component of the spectrogram, avoiding explicit feature extraction. In the proposed method, we use both the magnitude and the phase component. Contrary to~\cite{Chakrabarty2017}, which employed omnidirectional sensors only, general arrays with directional microphones additionally encode the DOA information in magnitude differences, while Ambisonics format especially encode directional information mainly in the magnitude component. f) All previous methods were evaluated on speech recordings that were synthetically spatialized and spatially static. We continue to use the static sound sources in the present work and extend them to a larger variety of sound events, such as impulsive and transient sounds. 


\section{Method}
\label{sec:method}
\begin{figure}[!tb]
 \centering 
\centerline{\includegraphics[width=\columnwidth, keepaspectratio]{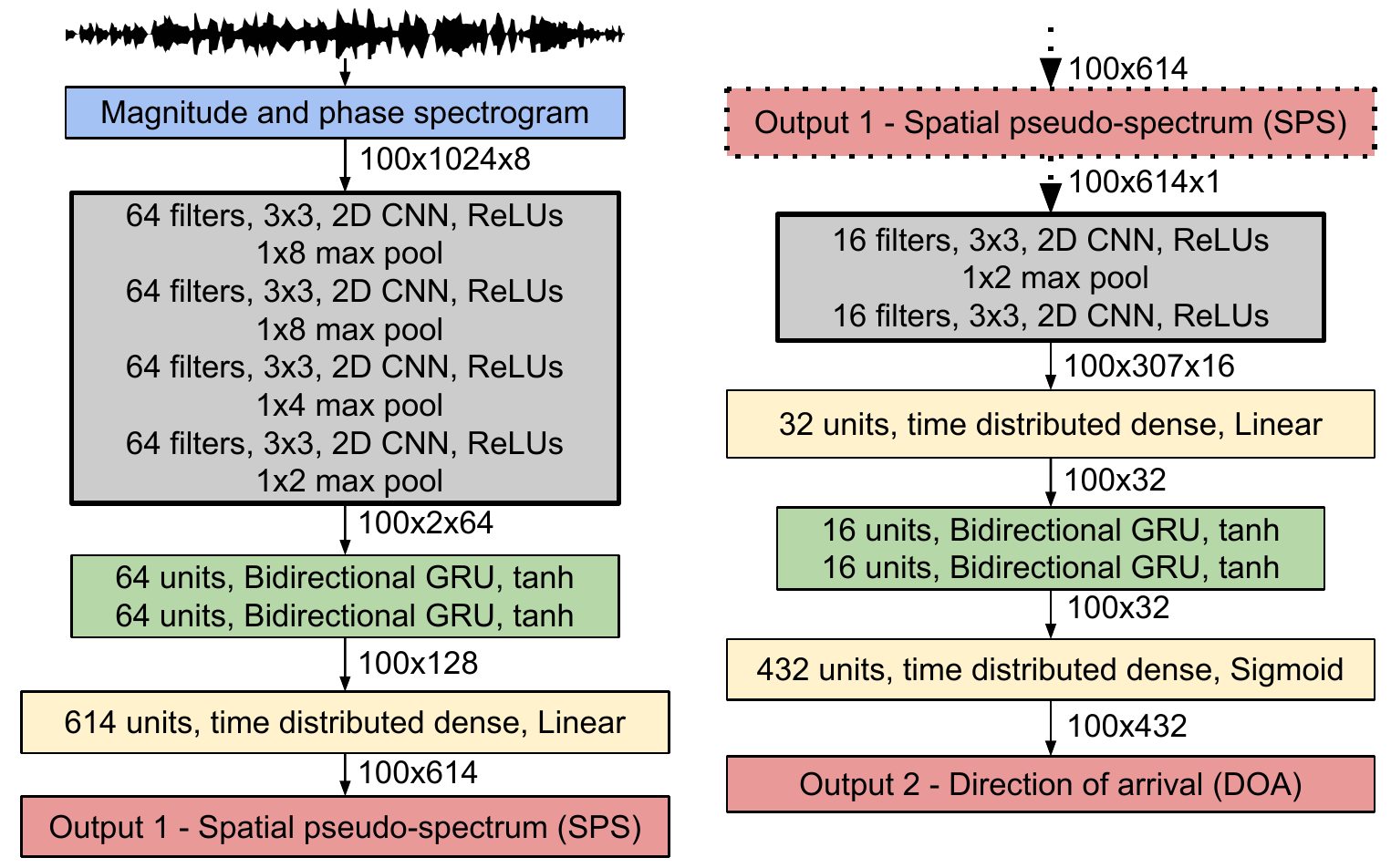}}
  	\caption{DOAnet - neural network architecture for direction of arrival estimation of multiple sound sources.}
  	\label{fig:crnn}    
    \vspace{-15pt}
\end{figure}

The block diagram of the proposed DOAnet is presented in Figure~\ref{fig:crnn}. The DOAnet takes multichannel audio as the input and first extracts the spectrograms of all the channels. The phases and the magnitudes of the spectrograms are mapped using a CRNN to two outputs sequentially. The first output, spatial pseudo-spectrum (SPS) is generated as a regression task, followed by the DOA estimates as a classification task. The DOA is defined by the azimuth $\phi$ and elevation $\lambda$ with respect to the microphone and the SPS is the intensity of sound along the DOA given by $S(\phi, \lambda)$. 

In this paper, we use discrete $\phi$ and $\lambda$ by uniformly sampling the 2-D polar coordinate space, with a resolution of 10 degrees in both azimuth and elevation, resulting in 614 sampled directions. The SPS is computed at each sampled direction, whereas, a subset of 432 directions is used for DOA, where the elevations are limited between -60 and 60 degrees.


\subsection{Feature extraction} The spectrogram is calculated for each of the audio channels whose sampling frequencies are 44100 Hz. A 2048-point discrete Fourier transform (DFT) is calculated on Hamming windows of 40 ms with 50~\% overlap. We keep 1024 values of the DFT corresponding to the positive frequencies, without the zeroth bin. $L$ frames of features, each containing 1024 magnitude and phase values of the DFT extracted in all the $C$ channels, are stacked in a $L\times 1024 \times 2C$ 3-D tensor and used as the input to the proposed neural network. The $2C$ dimension results from ordering the magnitude component of all channels first, followed by the phase. We use a sequence length $L$ of 100 (= 2 s) in this work.

\subsection{Direction of arrival estimation network (DOAnet)} Local shift-invariant features are extracted from the input spectrogram tensor ($L\times 1024 \times 2C$ dimension) using CNN layers. In every CNN layer, the intra-channel time-frequency features are processed using a receptive field of $3\times3$, rectified linear unit (ReLU) activation and pad zeros to the resulting activation map to keep the output dimension equal to input. Batch normalization and max-pooling operation along frequency axis are performed after every CNN layer to reduce the final dimension to $L\times 2 \times N_{C}$, where $N_{C}$ is the number of CNN filters in the last CNN layer. The CNN activations are reshaped to $L\times2N_{C}$ keeping the time axis length unchanged and fed to RNN layers in order to learn temporal structure. Specifically, the bi-directional gated recurrent units (GRU) with tanh activation are used. Further, the RNN output is mapped to the first output, the SPS, in regression manner using FC layers with linear activation.

The SPS is further mapped to DOA estimates--the final output of the proposed method--using a similar CRNN network as above with two minor architectural changes. An FC layer is introduced between the CNN and RNN layers to reduce the dimension of the RNN output. Additionally, the output layer which predicts the DOA uses sigmoid activation in order to estimate more than one DOA for a given time frame. Each node in this output layer represents a direction in 2-D polar space. During testing, the probabilities at these nodes are thresholded with a value of 0.5, so that anything greater suggests the presence of a source in the direction or otherwise absence of source.

We refer to the combined architecture of  SPS and DOA estimation in this work as DOAnet. The DOAnet is trained using the target SPS computed at each sampled direction, and for every time frame applying MUSIC (see Section~\ref{ssec:baseline}), and is represented using nonnegative real numbers. For the DOA output, the DOAnet aims to make a discrete decision about the presence of a source in a certain direction; and during training, the DOAnet uses the ground truth DOAs utilized to synthesize the audio (see Section~\ref{ssec:dataset}).  

The DOAnet was trained for 1000 epochs using Adam optimizer, mean squared error loss for SPS output and binary cross entropy loss for DOA output. The sum of the two losses was used for back propagation. Dropout was used after every layer and early stopping was used if the DOA metric (Section~\ref{ssec:metric}) did not improve for 100 epochs. The DOAnet was implemented using Keras framework with Theano backend.
 
\section{Evaluation}
\label{sec:eval}

\subsection{Dataset} \label{ssec:dataset}
In order to evaluate the proposed DOAnet, there are no publicly available real or synthetic datasets which consist of general sound events each associated with a 2D spatial coordinate. Since DNN-based methods need sufficiently large datasets to train on, most DNN-based methods proposed~\cite{Roden2015, Xiao2015, Zermini2016, Vesperini2016, Chakrabarty2017} have studied the performance on synthetic datasets. In similar fashion, we evaluate the proposed DOAnet on synthetic datasets about the same size as in the previous works.
 
We synthesize datasets consisting of static point sources associated with a spatial coordinate in the space in two contexts - anechoic and reverberant. For each context, three datasets are generated with no temporally overlapping sources ($O1$), maximum two overlapping sources ($O2$), and maximum three overlapping sound sources ($O3$). We refer to the anechoic context dataset as $OxA$ and reverberant as $OxR$, where $x$ denotes the number of overlapping sources. Each of these datasets has three cross-validation (CV) splits with 240 recordings for training and 60 for testing. Recordings are sampled at 44.1 kHz and 30 s long. 

In order to generate these datasets, we use the isolated real-life sound event recordings from the DCASE 2016 task 2~\cite{dcase2016Task2}. This dataset consists of 11 sound event classes, each with 20 examples. The classes in this dataset included speech, coughing, door slam, page-turning, phone ringing and keyboard sounds. During CV, for each of the splits, we randomly chose disjoint sets of 16 and 4 examples for training and testing, amounting to 176 examples for training and 44 for testing. In order to synthesize a recording, a random subset of the 176 or 44 sound examples was chosen from the respective split. The subset size varied for each recording based on the chosen sound examples. We start synthesizing a recording by randomly choosing the beginning time of the first randomly chosen sound example within the first second of the recording. The next randomly chosen sound example is placed 250-500 ms after the end of the first sound example. On reaching the maximum recording length of 30 s, the process is repeated as many times as the number of required overlapping sound events.

Each of the sound examples were assigned a DOA randomly using the following conditions. All sound events were placed in a spatial grid of ten degrees resolution along both azimuth and elevation. Two temporally overlapping sound events have at least ten degrees of spatial separation to avoid spatial overlapping. The elevation was constrained within the range of [-60, 60] degrees, as most natural sound events occur in this range. Finally, for the anechoic dataset, the sound sources were randomly placed at a distance $d$ in the range 1-10 m. For the reverberant dataset, the sound events were randomly placed inside a room of dimensions $10\times8\times4$ m with the microphone in the center of the room.

Spatialization for the anechoic case was done as following. Each point source signal $s_i$ with DOA $(\phi_i,\lambda_i)$, was converted to Ambisonics format by multiplying the signal with the vector $\mathbf{y}(\phi_i,\lambda_i) = [Y_{00}(\phi_i,\lambda_i), Y_{1(-1)}(\phi_i,\lambda_i), Y_{10}(\phi_i,\lambda_i), Y_{11}(\phi_i,\lambda_i)]^\mathrm{T}$ of real orthonormalized spherical harmonics $Y_{nm}(\phi,\lambda)$.  The complete anechoic sound scene multichannel recording $\mathbf{x}_\mathrm{A}$ was generated as $\mathbf{x}_\mathrm{A} = \sum_{i}g_i s_i \mathbf{y}(\phi_i,\lambda_i)$, with the gains $g_i<1$ modeling the distance attenuation. Each entry of $\mathbf{x}_\mathrm{A}$ corresponds to one channel and $g_i = \sqrt{1 / 10^{d/d_{max}}}$, where $d_{max} = 10$ m is the maximum distance.

In the reverberant case, a fast geometrical acoustics simulator was used to model natural reverberation based on the rectangular room image-source model \cite{Allen1979}. For each point source $s_i$ with DOA in the dataset, $K$ image sources were generated modeling reflections up to a predefined time-limit. Based on the room and its propagation properties, each image source was associated with a propagation filter $h_{ik}$ and DOA  $(\phi_k,\lambda_k)$ resulting in the spatial impulse response $\mathbf{h}_i = \sum_{k=1}^K h_{ik} \mathbf{y}(\phi_k,\lambda_k).$ The reverberant scene signal was finally generated by $\mathbf{x}_\mathrm{R} = \sum_{i} s_i*\mathbf{h}_i,$ where $(*)$ denotes convolution of the source signal with the spatial impulse responses. The room absorption properties were adjusted to match reverberation times of typical office spaces. Three sets of testing data were generated with similar room size as training data (Room 1), $80\%$ of room size ($8\times8\times4$ m) and reverberation time (Room 2), and $60\%$ of room size ($8\times6\times4$ m) and reverberation time (Room 3).

\subsection{Baseline}
\label{ssec:baseline}
The proposed method to our knowledge is the first DNN-based implementation for 2D DOA estimation of multiple overlapping sound events. Thus in order to evaluate the complete features of the proposed DOAnet, we compare the performance with the conventional, high-resolution DOA estimator based on MUSIC. Similar to the SPS and DOA outputs estimated by the DOAnet, the MUSIC method also estimates SPS and DOA, thus allowing a direct one-to-one comparison. 

The MUSIC SPS is based on a measure of orthogonality between the signal subspace (dominated by the source signals) of the spatial covariance matrix $\mathbf{C_s}$ and the noise subspace (dominated by diffuse and ambient sounds, late reverberation, and microphone noise). The spatial covariance matrix is calculated as $\mathbf{C_s} = \mathbb{E}_{f,t}\left[\mathbf{X}(f,t)\mathbf{X}(f,t)^\mathrm{H}\right]$, where spectrogram $\mathbf{X}(f,t)$ is a frequency $f$ and time $t$ dependent $C$-dimensional vector, where $C$ is the number of channels, $^\mathrm{H}$ is the conjugate transpose and $\mathbb{E}_{f,t}$ denotes the expectation over $f$ and $t$. For a sound scene with $O$ number of sources, the MUSIC SPS $S_{GT}$ is obtained from $\mathbf{C_s}$ by first performing an eigenvalue decomposition on $\mathbf{C_s} = \mathbf{E}\mathbf{\Lambda}\mathbf{E}^\mathbf{H}$. The sorted eigenvectors $\mathbf{E}$  (according to eigenvalues with decreasing magnitude) are further partitioned into the two aforementioned subspaces $\mathbf{E} = [\mathbf{U}_\mathrm{s}\; \mathbf{U}_n]$, where  $\mathbf{U}_\mathrm{s}$ denotes the signal subspace and will be composed of $O$ eigenvectors corresponding to the higher eigenvalues and the rest will form the noise subspace $\mathbf{U}_\mathrm{n}$. The $S_{GT}$ along the direction $(\phi_i,\lambda_i)$ is now given by $S_{GT}(\phi_i,\lambda_i) = 1/(\mathbf{y}^\mathrm{T}(\phi_i,\lambda_i)\mathbf{U}_n \mathbf{U}_n^\mathrm{H}\mathbf{y}(\phi_i,\lambda_i))$. Finally, the source DOAs are found by selecting the directions $(\phi_i,\lambda_i)$ corresponding to the $O$ largest peaks from $S_{GT}$.

\subsection{Metric}
\label{ssec:metric}
The DOAnet estimated SPS ($S_{E}(\phi, \lambda)$) is evaluated with respect to the baseline MUSIC estimated ground truth ($S_{GT}(\phi, \lambda)$) using the SNR metric calculated as $SNR = 10 \log_{10} (\sum_{\phi}\sum_{\lambda}S_{GT}(\phi, \lambda)^2 / \sum_{\phi}\sum_{\lambda} (S_{E}(\phi, \lambda) - S_{GT}(\phi, \lambda))^2)$.

As the DOA metric we use the angle between the estimate DOA (defined by azimuth $\phi_E$ and elevation $\lambda_E$) and the ground truth DOA ($\phi_{GT}$, $\lambda_{GT}$) used to synthesize the dataset in degrees. This is calculated as $\sigma = \arccos (\sin\phi_E\sin\phi_{GT} + \cos\phi_E\cos\phi_{GT} \cos(\lambda_{GT}-\lambda_E)) \cdot 180.0 / \pi$. Further, to accommodate the scenario of unequal number of estimated and ground truth DOAs we calculate and report the minimum distance between them using the Hungarian algorithm~\cite{Hungarian} along with the percentage of frames in which the number of DOAs estimated were correct. The final metric for the entire dataset, referred as DOA error, is calculated by normalizing the minimum distance with the total number of estimated DOA's.

\begin{figure}[!b]
\vspace{-15pt}
\begin{minipage}[b]{0.48\linewidth}
  \centering
  \centerline{\includegraphics[height=4cm,width=\columnwidth,trim={1.5cm 3cm 2cm 3cm},clip,keepaspectratio]{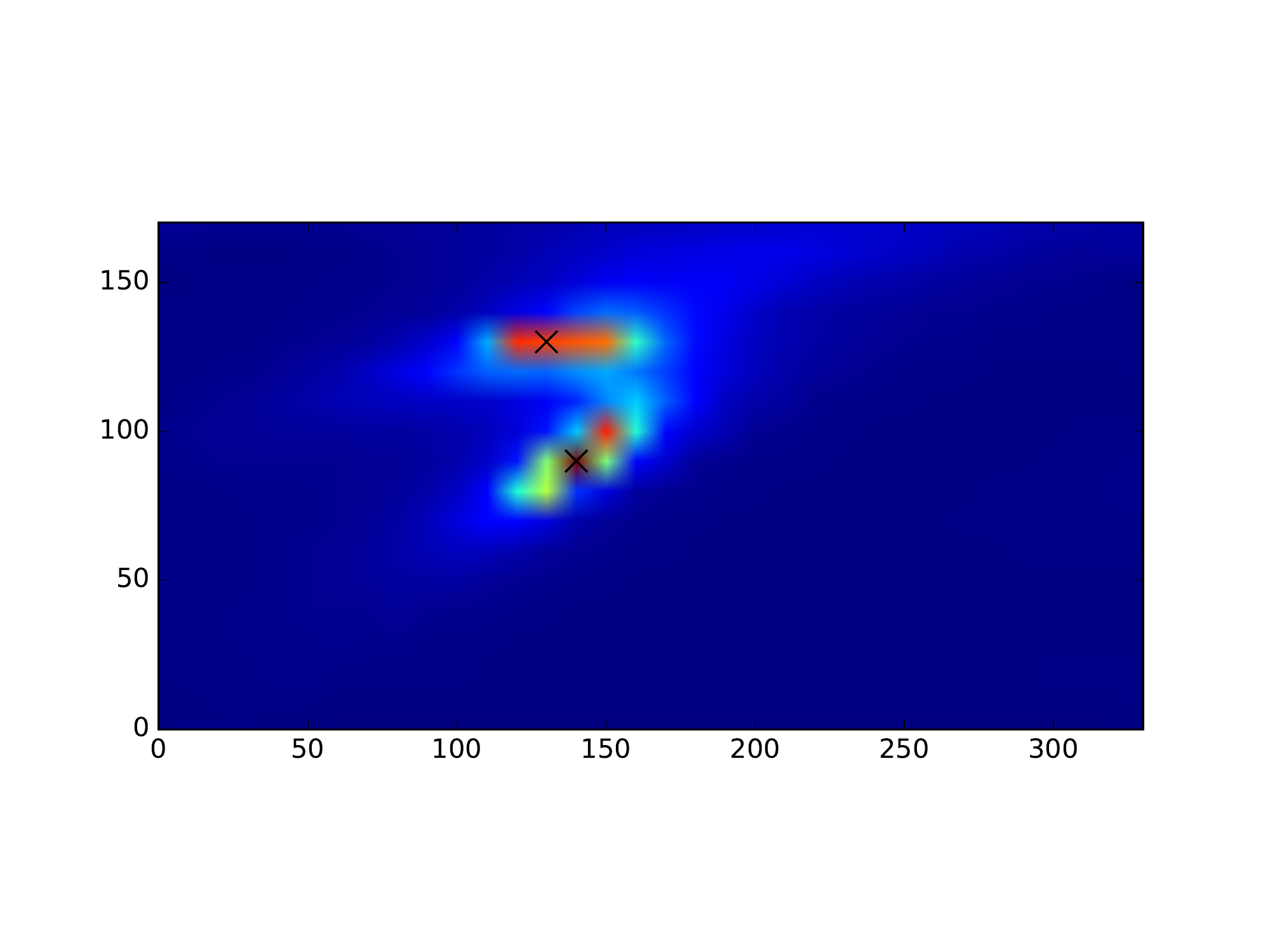}}
  \centerline{(a) MUSIC estimated }\medskip
\end{minipage}
\begin{minipage}[b]{0.48\linewidth}
  \centering
  \centerline{\includegraphics[height=4cm,width=\columnwidth, trim={1.5cm 3cm 2cm 3cm},clip,keepaspectratio]{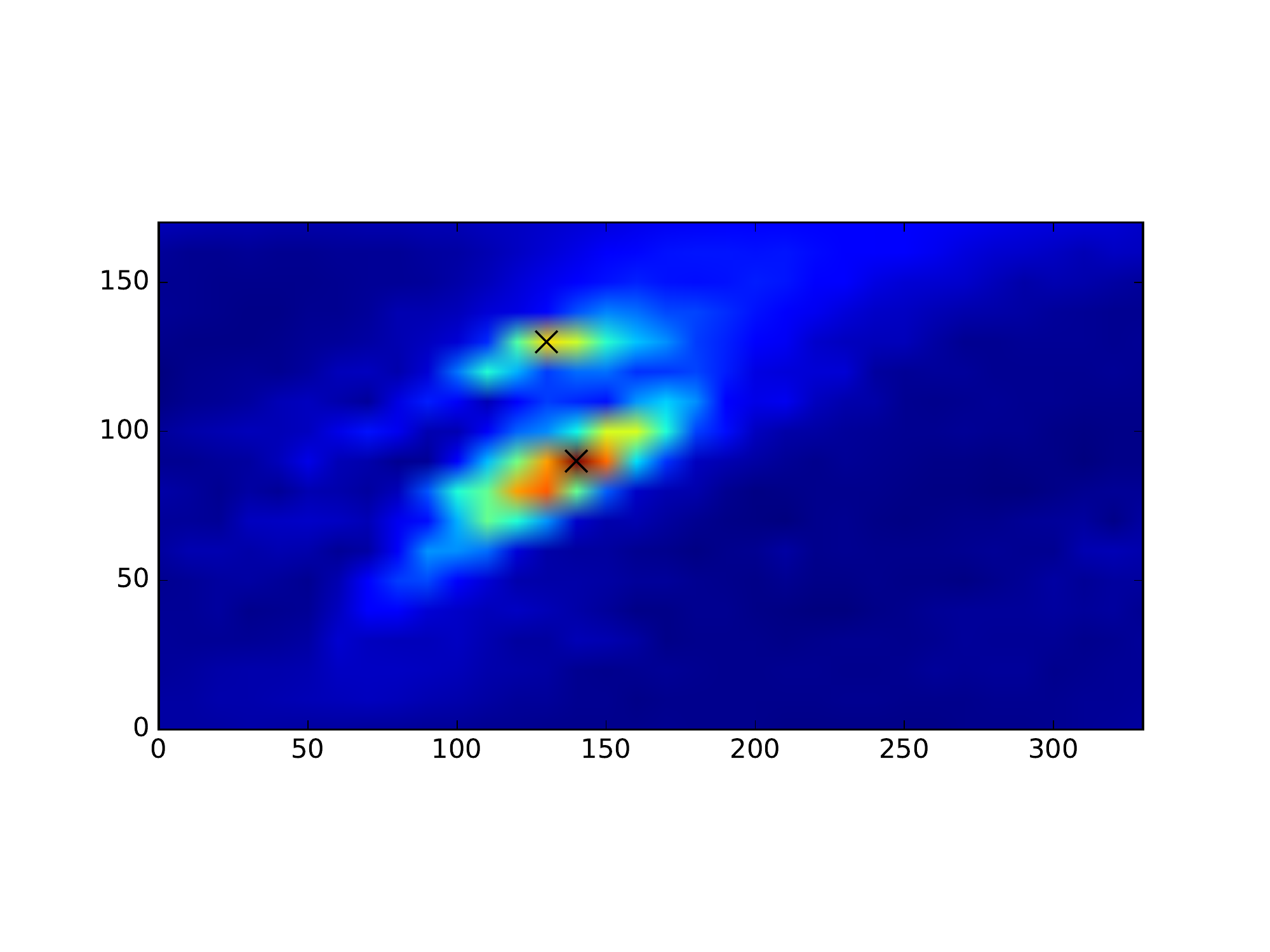}}
\centerline{(b) DOAnet estimated}\medskip
\end{minipage}
\vspace{-15pt}
\caption{SPS for two closely located sound sources. The black-cross markers represent the ground truth DOA. The horizontal axis is azimuth and vertical axis is elevation angle (in degrees)}
\label{fig:spm}
\end{figure}

\subsection{Evaluation procedure}
The parameter tuning for DOAnet was performed on the $O1A$ test data, and the best configuration is as shown in Figure~\ref{fig:crnn}. This configuration has 677 K weights, and the same configuration is used in all of the following studies.

At test time, the SNR metric for SPS output of the DOAnet ($S_{E}$) is calculated with respect to SPS of baseline MUSIC ($S_{GT}$). The DOA metric for the DOAs predicted by DOAnet and baseline MUSIC are calculated with respect to the ground truth DOA used to synthesize the dataset.

In the above experiment, the baseline MUSIC algorithm uses the knowledge of the number of active sources. In order to have a fair evaluation, we test the DOAnet in a similar scenario where the number of sources is known. We use this knowledge to choose the top probabilities in prediction layer of the DOAnet instead of thresholding it with a value of 0.5. 

\begin{table}[!tb]
\caption{Evaluation metric scores for the spatial power map and DOAs estimated by the DOAnet for different datasets.}
\label{T2:results}
\resizebox{\columnwidth}{!}{
\centering
\begin{tabular}{l|ccc|ccc}
 & \multicolumn{3}{c|}{Anechoic} & \multicolumn{3}{c}{Reverberant (Room 1)} \\\hline

\begin{tabular}[c]{@{}l@{}}Max. no. of  \\ overlapping sources\end{tabular} & 1 & 2 & 3 & 1 & 2 & 3 \\ \hline
SPS SNR (in dB) & 9.90 & 3.35 & -0.26 & 3.11 & 1.24 & 0.13 \\ 
\multicolumn{7}{l}{\multirow{2}{*}{}} \\
\multicolumn{7}{l}{DOA error with unknown number of active sources (threshold of 0.5)} \\\hline
DOAnet & 0.57 & 8.03 & 18.34 & 6.31 & 11.46 & 38.41 \\
\begin{tabular}[c]{@{}l@{}}Correctly predicted \\ frames (in \%)\end{tabular} & 95.4 & 42.7 & 1.8 & 59.3 & 15.8 & 1.2 \\
\multicolumn{7}{l}{\multirow{2}{*}{}} \\
\multicolumn{7}{l}{DOA error with known number of active sources} \\\hline
DOAnet & 1.14 & 27.52 & 49.30 & 12.61 & 38.98 & 67.07 \\
MUSIC & 2.29 & 8.60 & 28.66 & 25.80 & 57.33 & 91.72 
\end{tabular}
}
\vspace{-15pt}
\end{table}


\begin{figure}[!hb]
\vspace{-15pt}
\hspace{11pt}
\begin{minipage}[b]{0.10\linewidth}
  \centering  
  \centerline{\includegraphics[height=2cm,trim={8cm 2.5cm 8cm 3.5cm},clip,keepaspectratio]{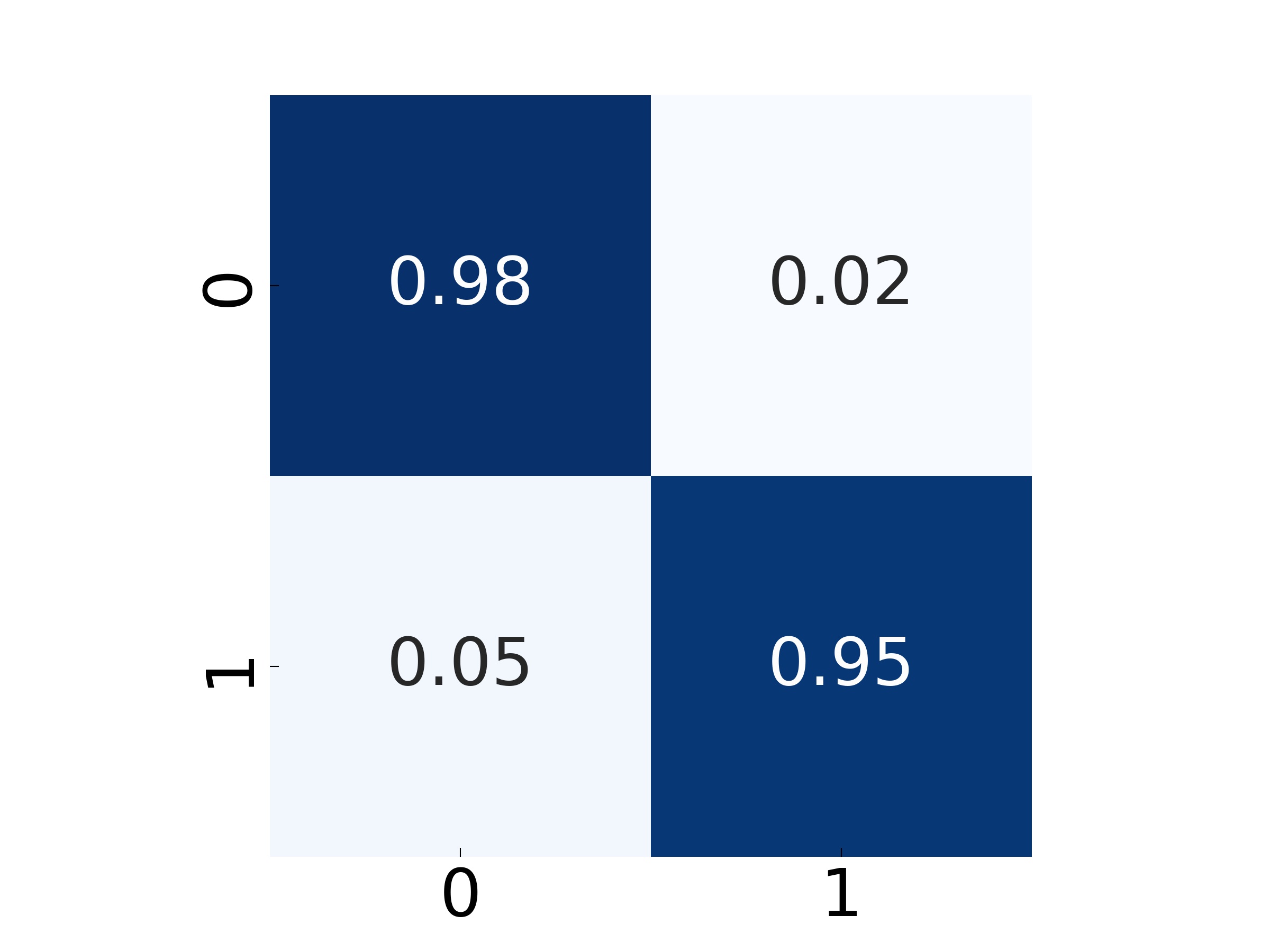}} 
  \centerline{(a) $O1A$}\medskip
\end{minipage}
\hspace{5pt}
\begin{minipage}[b]{0.35\linewidth}
  \centering  
  \centerline{\includegraphics[height=2cm, trim={6cm 2.5cm 8cm 3.5cm},clip,keepaspectratio]{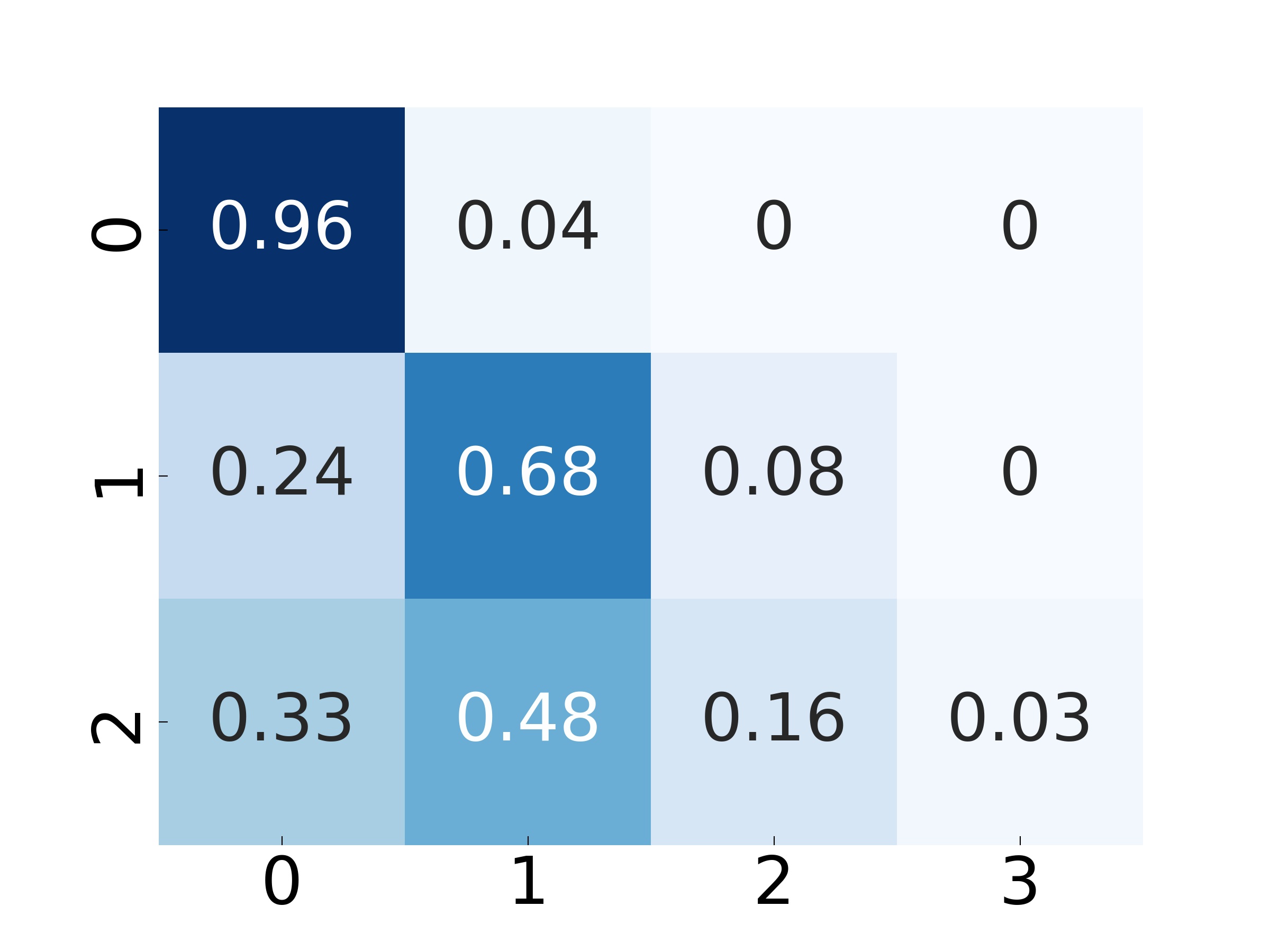}}  
  \centerline{(b) $O2A$}\medskip
\end{minipage}
\hspace{-10pt}
\begin{minipage}[b]{0.25\linewidth}
  \centering
  \centerline{\includegraphics[height=2cm, trim={8.5cm 2.5cm 8cm 3.5cm},clip,keepaspectratio]{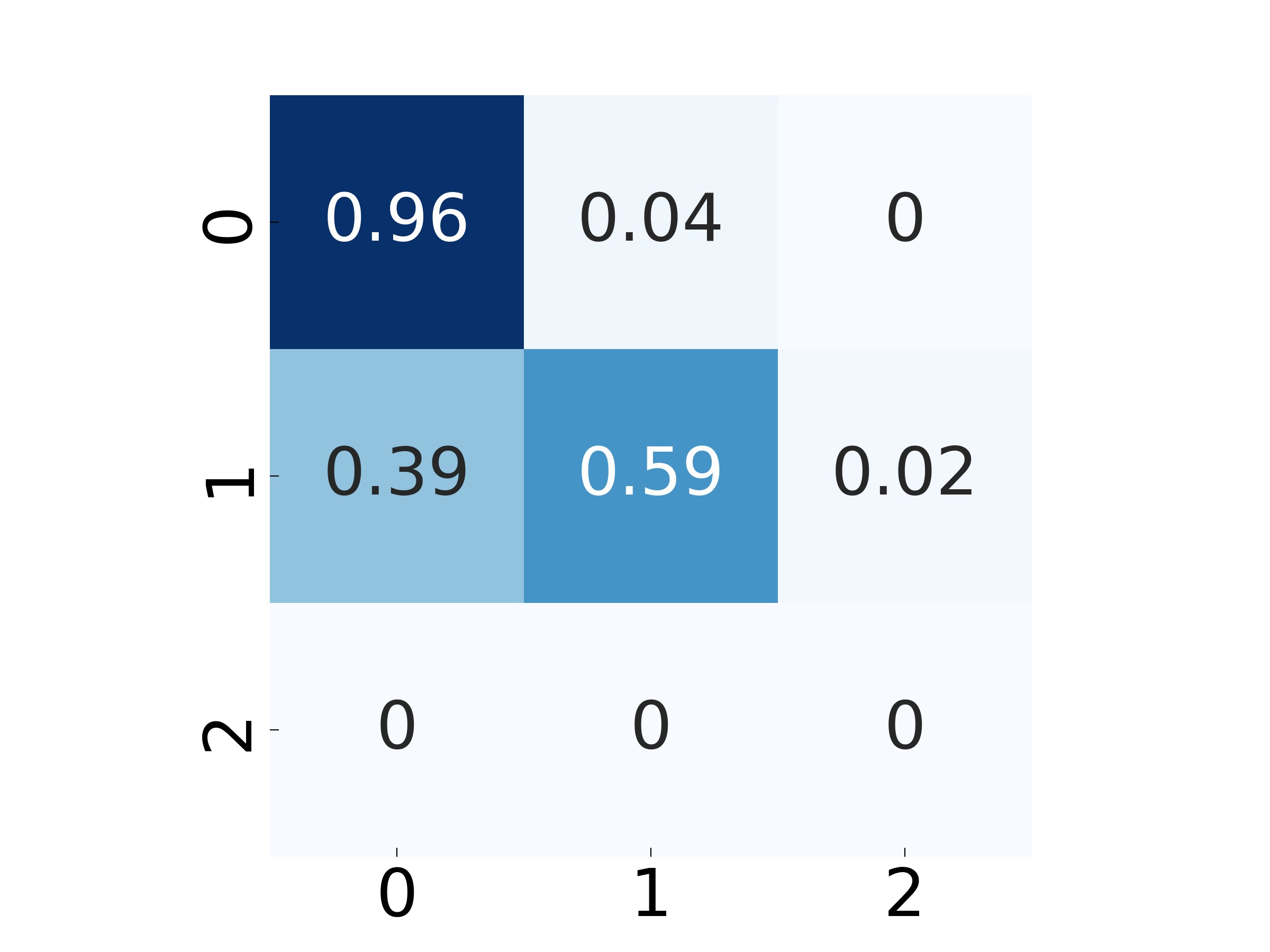}}
  \centerline{(c) $O1R$}\medskip
\end{minipage}
\hspace{-10pt}
\begin{minipage}[b]{0.25\linewidth}
  \centering  
  \centerline{\includegraphics[height=2cm, trim={8.5cm 2.5cm 8cm 3.5cm},clip,keepaspectratio]{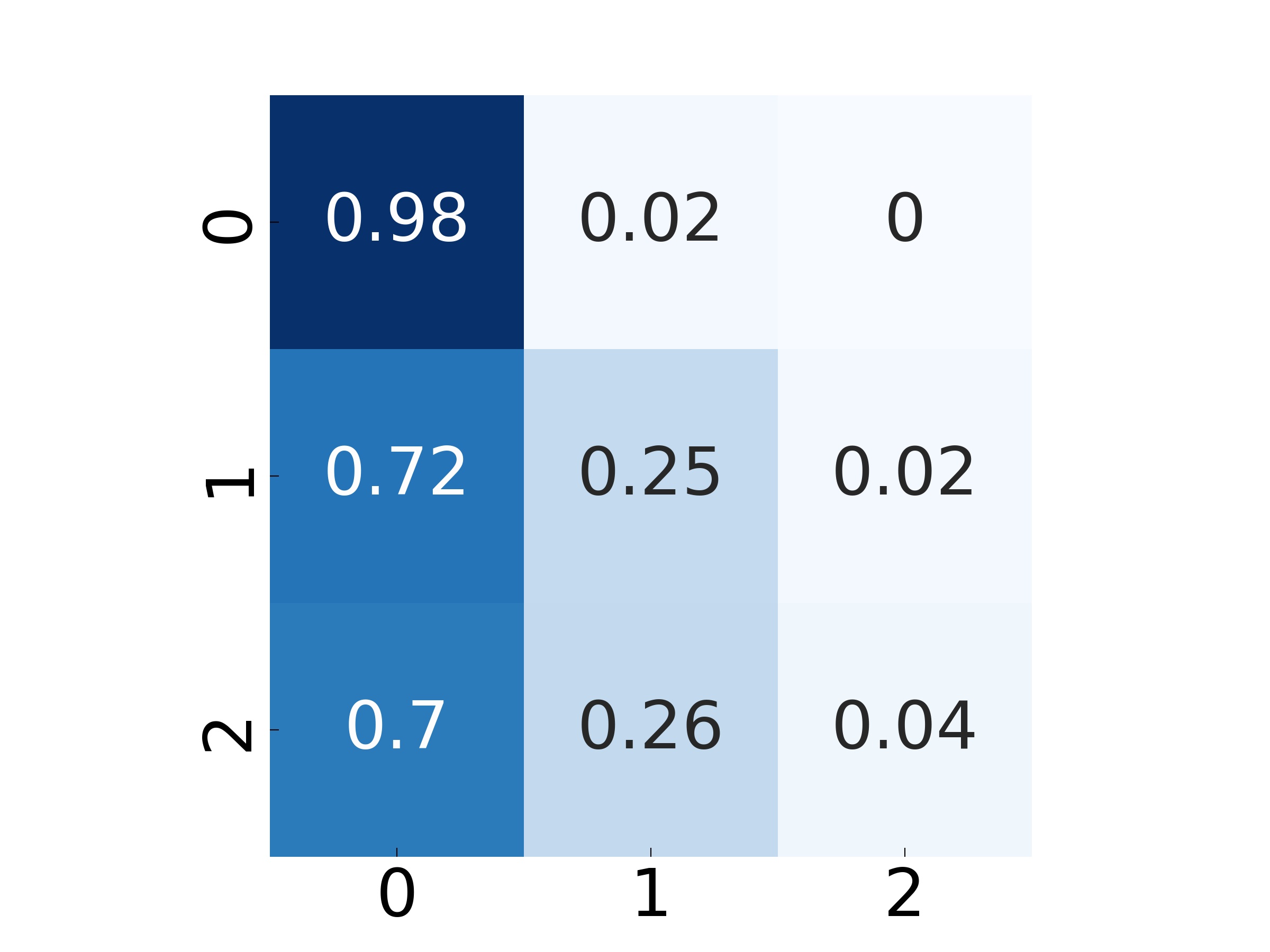}}
  \centerline{(d) $O2R$}\medskip
\end{minipage}
\vspace{-25pt}
\caption{Confusion matrix for the number of DOA estimated per frame by the DOAnet. The horizontal axis is the DOAnet estimate, and the vertical axis is the ground truth.}
\label{fig:conf_mat}
\end{figure}

\section{Results and Discussion}
\label{sec:results}
The results of the evaluations are presented in Table~\ref{T2:results}. The high SNRs for SPS in both the contexts, with up to one and two overlapping sound events show that the SPS generated by DOAnet ($S_{E}$) is comparable with the baseline MUSIC SPS ($S_{GT}$). Figure~\ref{fig:spm} shows the $S_{E}$ and the respective $S_{GT}$ when two active sources are closely located. In the case of up to three overlapping sound events, the baseline MUSIC is already at its theoretical limit of estimating $N-1$ sources from $N$-dimensional signal space~\cite{Ottersten1993}. In practice, for $N-1$ sources only one noise subspace vector $\mathbf{U}_n$ is used to generate SPS, which for real signals is too weak for stable estimation. In the present evaluation of DOAnet which is trained with four-channel audio features and MUSIC SPS, for the case of three overlapping sound sources the SPS used is an unstable estimate resulting in poor training and consequently the results. With more than four-channels input, which the proposed DOAnet can easily extend to, it can potentially localize more than two sound sources simultaneously.

The DOA error for the proposed DOAnet when the number of active sources are unknown is presented in Table~\ref{T2:results}. The DOAnet error is considerably better in comparison to the baseline MUSIC that uses the active sources knowledge for all datasets. However, the number of frames in which DOAnet produced the correct number of active sources were few. For example, in the case of anechoic recordings with up to two overlapping sound events, only 42.7\% of the estimated frames had the correct number of DOA predictions. This prediction drops even drastically when the number of sources is three, due to the theoretical limit of MUSIC as explained previously, and consequently for the DOAnet as MUSIC SPS is used for training. Finally, the confusion matrix for the number of DOA estimates per frame for $O1$ and $O2$ datasets are visualized in Figure~\ref{fig:conf_mat}. We skipped the confusion matrices for the $O3$ datasets as they were not meaningful for similar reasons as explained above.

With the knowledge of the number of active sources (Table~\ref{T2:results}), the DOAnet performs considerably better than baseline MUSIC for all datasets other than the $O2A$ and $O3A$. The MUSIC DOA's were chosen using a 2D peak finder on the MUSIC SPS, whereas the DOA's in DOAnet were chosen by simply picking the top probabilities in the final DOA prediction layer. A smarter peak picking method from the DOAnet, or using the number of sources as an additional input can potentially result in better scores across all datasets. Further, the DOAnet error on unmatched reverberant data is presented in Table~\ref{T3:reverb}. The performance of DOAnet is seen to be consistent in comparison to the matched reverberant data in Table~\ref{T2:results}, and significantly better than the performance of MUSIC.

In this paper, since the baseline was chosen to be MUSIC, for a fair comparison the DOAnet was also trained using MUSIC SPS. In an ideal scenario, considering the DOAnet is trained using datasets for which the ground truth DOAs are known, we can generate accurate high-resolution SPS from the ground truth DOA's as per the required application and use them for training. Alternatively, the DOAnet can be trained without the SPS to directly generate the DOAs, it was only used in this paper to present the complete potential of the method in the limited paper space. In general, the above results show that the proposed DOAnet has the potential to learn the 2D direction information of multiple overlapping sound sources directly from the spectrogram of the input audio without the knowledge of the number of active sound sources. An exhaustive study with more detailed experiments including both synthetic and real datasets are planned for future work.


\begin{table}[!t]
\caption{Evaluation scores for unmatched reverberant room.}
\label{T3:reverb}
\centering
\resizebox{0.9\columnwidth}{!}{
\begin{tabular}{l|cc|cc}
 & \multicolumn{2}{c|}{Room 2} & \multicolumn{2}{c}{Room 3} \\\hline
Max. no. of overlapping sources  & 1 & 2 & 1 & 2 \\ \hline
SPS SNR (in dB) & 3.53 & 1.49 & 3.49 & 1.46 \\
\multicolumn{5}{l}{\multirow{2}{*}{}} \\
\multicolumn{5}{l}{DOAnet error (Unknown number of sources)} \\\hline
DOAnet & 3.44 & 6.88 & 4.59 & 10.89\\
Correctly predicted frames (in \%) & 46.2 & 14.3 &  49.7 & 14.1 \\ 
\multicolumn{5}{l}{\multirow{2}{*}{}} \\
\multicolumn{5}{l}{DOA error (Known number of sources)} \\\hline
DOAnet & 8.60 & 32.10 & 9.17 & 33.82\\
MUSIC & 31.52 & 58.47 & 33.25 & 60.76
\end{tabular}
 } 
\vspace{-15pt}
\end{table}

\section{Conclusion}
\label{sec:conclusion}
 A convolutional recurrent neural network (DOAnet) was proposed for multiple source localization. The DOAnet was shown to learn the number of active sources directly from the input spectrogram, and estimate precise DOA in 2-D polar space. The method was evaluated on anechoic, matched and unmatched reverberant dataset. The proposed DOAnet performed considerably better than baseline MUSIC in most scenarios. Thereby showing the potential of DOAnet in learning highly computational algorithm without prior knowledge of the number of sources. 


\bibliographystyle{IEEEtran}
\bibliography{refs}

\end{document}